# Colossal electroresistance effect around room temperature in LuFe$_2$O$_4$

Chang-Hui Li, Xiang-Qun Zhang, Zhao-Hua Cheng, and Young Sun[*]

*Beijing National Laboratory for Condensed Matter Physics, Institute of Physics, Chinese Academy of Sciences, Beijing 100080, P. R. China*

Abstract

A colossal electroresistance effect is observed around room temperature in a transition metal oxide LuFe$_2$O$_4$. The measurements of resistance under various applied voltages as well as the highly nonlinear current-voltage characteristics reveal that a small electric field is able to drive the material from the insulating state to a metallic state. The threshold field at which the insulating-metallic transition occurs, decreases exponentially with increasing temperature. We interpret this transition as a consequence of the breakdown of the charge-ordered state triggered by applied electric field, which is supported by the dramatic dielectric response in a small electric field. This colossal electroresistance effect as well as the high dielectric tunability around room temperature in low applied fields makes LuFe$_2$O$_4$ a very promising material for many applications.

[*]Electronic mail: youngsun@aphy.iphy.ac.cn

Effect of electric field on the resistivity of a material has attracted considerable attention due to its rich physics as well as great potential in developing electronic devices.[1-3] A large change of resistivity upon applied electric field, usually termed as the electroresistance effect, has been observed in a variety of transition metal oxides such as perovskite manganites,[4,5] titanates,[6,7] and magnetite.[8] Although the underlying physical mechanism could be distinct from each other, these effects are believed to be intrinsic properties of the materials, in contrast to those extrinsic electroresistance observed in some metal-insulator-metal (MIM) structures where the interfacial effects play a dominating role.[1] In most cases, these intrinsic electroresistance effects happen at quite low temperatures and require high electric fields, which is an apparent drawback for practical applications. Here we show that an intrinsic colossal electroresistance effect occurs around room temperature in $LuFe_2O_4$.

The mixed-valence material $LuFe_2O_4$ belongs to the family of the rare earth-iron oxide, $RFe_2O_4$, where R stands for rare-earth elements from Dy to Lu and Y.[9] It has a rhombohedral crystal structure consisting of alternate stacking of triangular lattices of rare-earth elements, iron and oxygen ions. As the Fe ions have an average valence of 2.5+, an equal amount of $Fe^{2+}$ and $Fe^{3+}$ ions coexist at the same site in the triangular lattice. The strong magnetic interactions between localized Fe moments develop as a two-dimensional (2D) ferrimagnetic ordering below the Neel temperature $T_N \sim 250$ K.[10] The system also involves a charge interaction between $Fe^{2+}$ and $Fe^{3+}$ ions. Compared with the average Fe valence of 2.5+, $Fe^{2+}$ and $Fe^{3+}$ ions are considered as having an excess and a deficiency of half an electron, respectively. Thus the coulombic interactions between $Fe^{2+}$ and $Fe^{3+}$ are accompanied by a charge frustration on the triangular lattice. The competing interactions between frustrated charges are settled by a $\sqrt{3} \times \sqrt{3}$ superstructure of $Fe^{2+}$ and $Fe^{3+}$ arrangement, similar to the stable configuration of Ising spins in a triangular lattice. As a consequence, $LuFe_2O_4$ exhibits a three-dimensional (3D) charge ordered (CO) state below 330 K.[10-12] Above 330 K, the 3D charge ordering transforms into a two-dimensional ordering state. Spectacularly, the charge ordering in $LuFe_2O_4$ leads to a local electrical polarization because the centers of $Fe^{2+}$ and $Fe^{3+}$ do not coincide in the unit cell of the superstructure.[13,14] Such kind of ferroelectricity associated with charge ordering is termed as "electronic ferroelectricity",[15] which is in contrast to conventional ferroelectricity involving displacement of cation and anion pairs. The charge-ordered state in many systems could be melted by external forces such as magnetic field, electric field, and pressure, which generally leads to an insulating to metallic transition.[16] For example, both the charge-ordered perovskite Manganites[4] and magnetite $(Fe_3O_4)$[8] exhibit an electrically driven phase transition below the charge-ordering temperature $T_{CO}$. Considering the high $T_{CO}$ in $LuFe_2O_4$, one may expect a similar phase transition induced by electric field at relatively high temperatures.

Figure 1 shows the temperature dependence of resistance of $LuFe_2O_4$ under various applied voltages. The resistance was measured with a conventional four-probe configuration (see the inset of Fig. 1). In a low applied voltage (0.3 V), the resistance shows an insulating behavior with an apparent kink around 330 K where the 3D to 2D charge ordering transition occurs. In higher applied voltages, the resistance is nearly the same at low temperature. However, as temperature increases, an abrupt drop of resistance takes place, which results in a burst of current exceeding the current limit of 100 mA and makes the lower resistance undetectable. With increasing applied voltage, the abrupt change in resistance starts at lower temperature. These sharp resistance drops indicate that the applied voltages may induce an insulating-metallic transition at high temperatures.

Figure 2 shows the current-voltage (I-V) curves at various temperatures. These curves were measured by scanning pulsed voltage up to 80 V. To protect the circuit against a burst of big current, the current limit is set as 100 mA. At 240 K, the I-V curve is close to linearity up to 80 V. Above 260 K, each I-V curve deviates from the linearity and exhibits a big jump of current at a threshold voltage, $V_{th}$. As shown in the inset of Fig. 2, the threshold voltage decreases nearly exponentially with increasing temperature so that a small voltage of a few volts is able to cause the transition at 340 K. We have confirmed that the sharp transitions are not due to Joule heating. We monitored the temperature of the sample during the voltage scan and found that the temperature increase was less than 0.8 K. Moreover, we used pulsed voltages/currents in all measurements so that the Joule heating is limited to a tiny effect. These highly nonlinear I-V curves further prove the point that an insulating-metallic transition can be induced by applied electric fields in $LuFe_2O_4$.

In order to better evaluate the electrically driven insulating-metallic transition, we measured the I-V hysteresis loop at 300 K. As shown in Fig. 3(a), the I-V loop looks quite symmetric with positive and negative voltages. The initial ascending branch, marked with 1, gives a threshold voltage ~ 22 V at which the current suddenly jumps up and exceeds the 100 mA limit. When the voltage is scanned back, the current remains undetectable till ~ 10 V and drops fast thereafter, giving a large hysteresis. When the voltage is scanned to negative, the switching phenomenon repeats at nearly the same threshold voltages. The final branch with increasing voltage, marked with 2, is slightly different from the initial branch, showing a smaller threshold voltage.

Figure 3(b) shows the resistance of $LuFe_2O_4$ as a function of the cycle of applied voltage at 300 K. The resistance decreases continuously with increasing voltage, which suggests that the nonlinearity in the I-V curves starts well before the switching. At the threshold voltage, the resistance becomes undetectable because the current exceeds the limit of 100 mA. When the voltage is scanned back, the resistance becomes detectable only below ~10 V and does not return to the initial value at zero voltage. The resistance increases further as the voltage is scanned to negative and reach a maximum at ~ -10 V. Then it decreases smoothly with increasing negative voltage and becomes undetectable at the negative threshold voltage. When the voltage is scanned back from the negative to the positive, a symmetric variation was observed. The relative change of resistance ($R/R_0$) as a function of applied voltage at various temperatures is shown in Fig. 4. $R_0$ is the initial resistance at the lowest voltage. The relative ratio is nearly 100% at the threshold voltages as the resistance changes by several orders, which is really a colossal electroresistance effect. We note that the measurements were performed on a bulk sample with a 3.8 mm length. Thus, the threshold voltage (~ 22 V) at room temperature corresponds to a quite small electric field of tens V/cm, which is beneficial for practical applications.

The electrically driven phase transition observed in $LuFe_2O_4$ has some common features with that in the charged-ordered $Pr_{1-x}Ca_xMnO_3$ manganites[4] and magnetite $(Fe_3O_4)$[8], but occurs at much higher temperatures and requires much smaller electric fields. It has been long believed that the resistive switch in manganites is due to the melting of the CO state triggered by applied electric field.[4,16,17] Since $LuFe_2O_4$ is also a CO system, we think that a similar mechanism could apply. In the CO state, the charge carriers are initially localized at each atomic site due to the repulsive electron-electron interaction. A high enough electric field may cause the dielectric breakdown of the CO state, which immediately leads to a large number of mobile charge carriers and, consequently, a metallic state appears. The fact that the threshold voltage decreases exponentially with increasing temperature implies that thermally-assisted fluctuations make the melting of the CO state much easier. Therefore, with the benefit of a high $T_{CO}$ above room temperature in $LuFe_2O_4$, a small electric field is able to break down the CO state in the vicinity of room temperature and induces an insulating-metallic transition.

The breakdown of the CO state by electric field can be evidenced by the dielectric response measurements. Since the "electronic ferroelectricity" in $LuFe_2O_4$ is associated with charge ordering of $Fe^{2+}$ and $Fe^{3+}$ ions[13], the breakdown of the CO state would subsequently result in a suppression of the electrical polarization, which should be reflected in the dielectric response. Fig. 5 shows the dielectric constant of $LuFe_2O_4$ as a function of temperature in zero and a 50 V/cm bias electric field. Apparently, the dielectric constant is greatly suppressed in a small electric field, by more than 50% around room temperature. Similarly, a strong suppression of the low-frequency dielectric constant by bias electric field was recently observed in the CO state of $Pr_{0.5}Ca_{0.5}MnO_3$ which also exhibits a colossal electroreisstance effect at low temperature.[17] The authors considered it as an evidence for the melting of the polaron ordering. Therefore, the result of dielectric response supports the picture of breakdown of the CO state by electric field. We note that the large change of dielectric constant in a biased electric field, known as the dielectric tunability effect, has potential usefulness for many devices.[18] The details of the high dielectric tunability in $LuFe_2O_4$ can be found elsewhere.[19]

We also examined the influence of magnetic field on the resistance and I-V characteristics. Unlike perovskite manganites where magnetic field has a strong action,[4,5] we found that the magnetoresistance effect in $LuFe_2O_4$ is very small even in a 10 T magnetic field. It seems that the electronic properties of $LuFe_2O_4$ are very sensitive to electric field but quite dull to magnetic field. The physics underlying this peculiar feature deserves further studies. Combining the multiple virtues of room temperature and colossal effects in low applied fields, $LuFe_2O_4$ is not only a very promising material for many applications but also provides a playground for intriguing physics of strongly correlated electrons.

This work was supported by the National Key Basic Research Program of China (2007CB925003) and the National Natural Science Foundation of China (50721001).

Experiments

The polycrystalline samples of $LuFe_2O_4$ were prepared by a solid state reaction method. A stoichiometric mixture of high-purity $Lu_2O_3$ (99.99%), $Fe_2O_3$ (99.998%), and Fe (99.99%) metal powders was well ground and pressed into pellets. The samples were sintered at 1100 °C in evacuated quartz tubes for 48 h. Powder X–ray diffraction at room temperature showed that the samples are single phase with a structure consistent with literature.

The resistance and I-V measurements were performed using the circuit shown in the inset of Fig. 1. The applied voltage was supplied by a Keithley 2400 source meter up to 80 V. To protect the circuit against a burst of big current, the current limit is set to 100 mA. The sample has a size of 3.8×1×1 (length× width×thickness) $mm^3$. An electrode was made at each end of the sample using silver paint. The voltage drop across the sample was monitored using two other electrodes attached to the surface with a space of 1 mm. The resistance was thus measured using the conventional four-probe configuration. The sample for dielectric measurements was prepared by applying silver electrodes to the polished surfaces of a thin pellet with a thickness of 1.0 mm and attaching an electrode to each face using silver paint. The dielectric constant was measured using a NF ZM2353 LCR meter. Direct current bias voltage was supplied by a Keithley 2400 source meter. The temperature during all the measurements was controlled by a Quantum Design Superconducting Quantum Interference Device.

Figure 1. Temperature dependence of resistance of $LuFe_2O_4$ measured in various applied voltages. The measurements were performed using the circuit shown in the inset. To protect the circuit against a burst of current, the current limit is set to 100 mA.

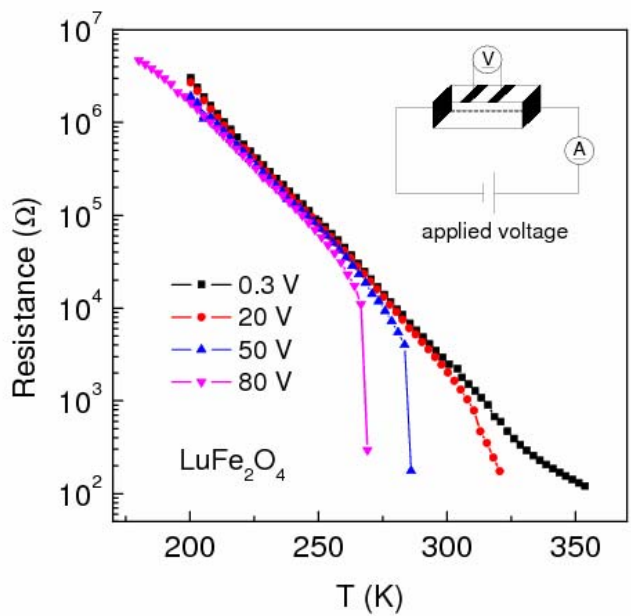

Figure 2. Current-voltage (I-V) curves at various temperatures. The measurements were performed by scanning pulsed voltage up to 80 V using the circuit shown in Fig. 1. The current limit is set to 100 mA in order to protect the circuit against a burst of big current. The sudden jump of current at a threshold voltage Vth indicates an insulating-metallic transition. The inset shows the temperature dependence of Vth.

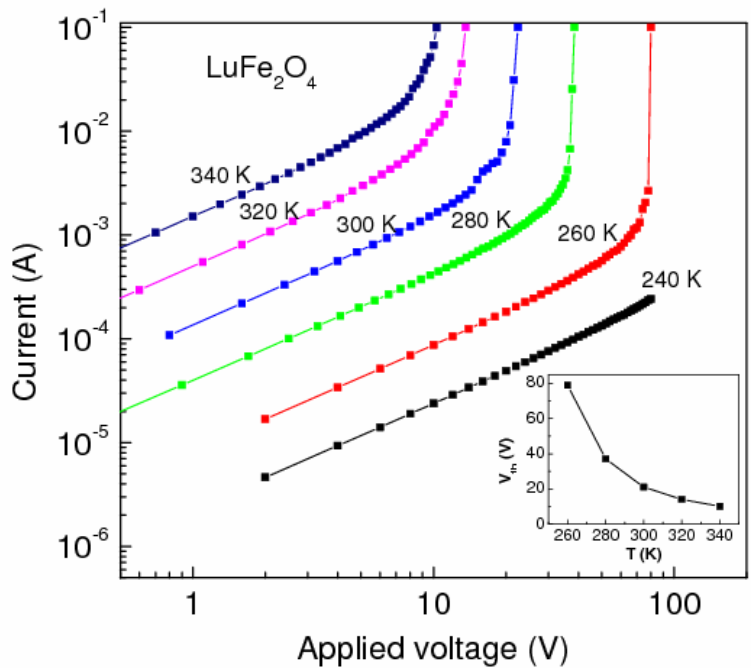

Figure 3. (a) The I-V hysteresis loop at 300 K. The initial branch with increasing voltage is marked with 1 and the final branch with increasing voltage is marked with 2. The solid squares linked with solid lines refer to the undetectable ranges where the current exceeds the limit of 100 mA. (b) The resistance as a function of the cycle of applied voltage at 300 K. The open squares linked with dot lines refer to the undetectable ranges where the current exceeds the limit of 100 mA.

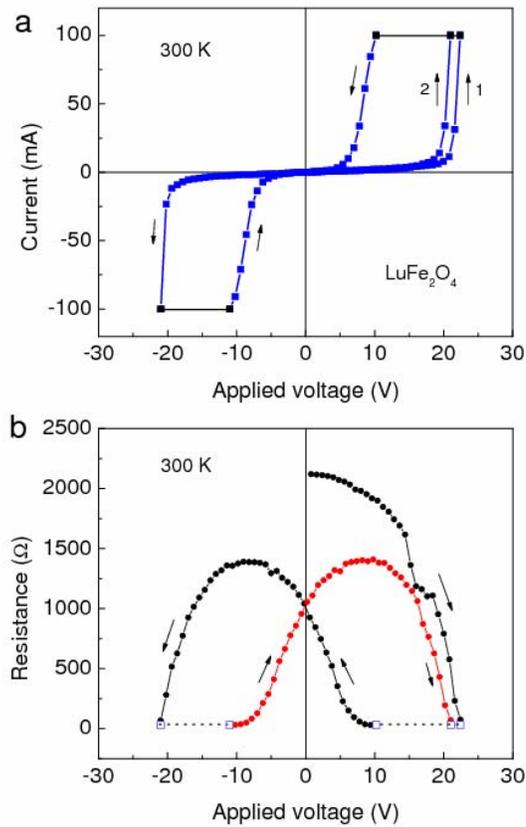

Figure 4. The relative change of resistance ($R/R_0$) as a function of applied voltage at various temperatures. $R_0$ is the initial resistance at the lowest voltage. For clarity, the voltage is plotted in a logarithmic scale.

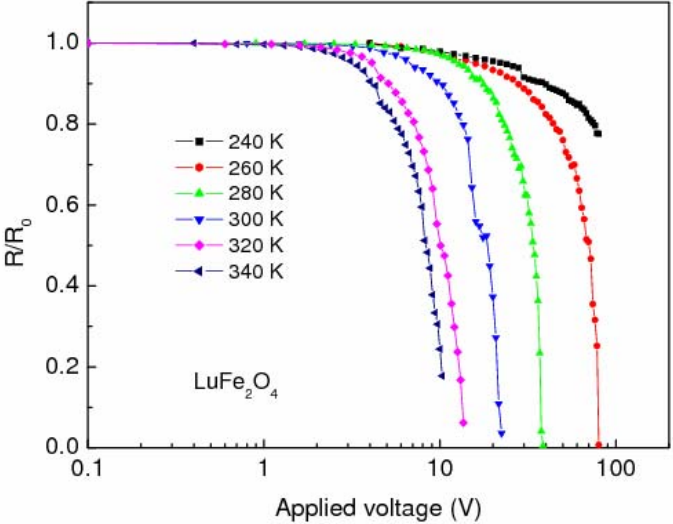

Figure 5. Influence of dc bias electric field on the dielectric constant of $LuFe_2O_4$. The dielectric constant is strongly suppressed by a small electric field of 50 V/cm, which implies the breakdown of the charge ordered state by applied electric field.

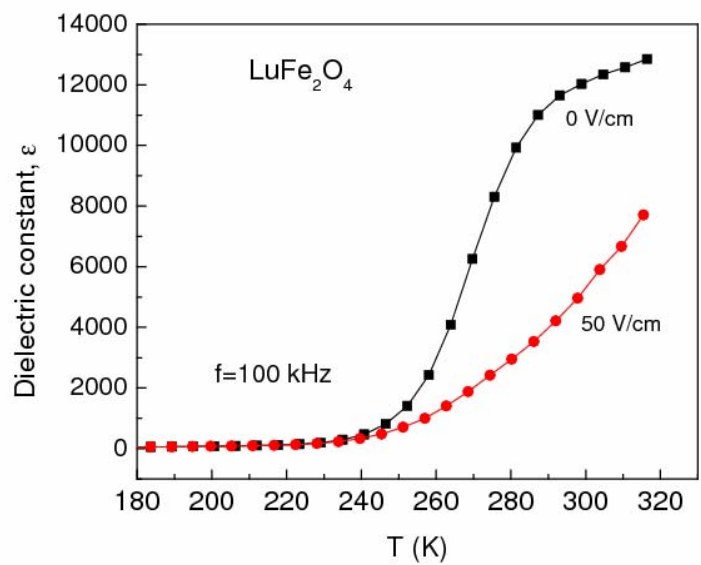